\documentclass[proceedings]{stacs}
\stacsheading{2009}{697--708}{Freiburg}
\firstpageno{697}

\usepackage{amssymb}

\def\zo{\{0,1\}}
\newcommand{\ie}{$\mbox{i.e.}$}
\def\mapping{\rightarrow}
\newcommand{\nat}{{\mathbb N}}
\newcommand{\real}{{\mathbb R}}
\newcommand{\rest}{{\upharpoonright}}

\newcommand{\zon}{\zo^n}

\newcommand{\dimm}{\rm dim}

\newcommand{\barx}{\overline{x}}
\newcommand{\bary}{\overline{y}}
\newcommand{\barxy}{\barx_{i-1}\bary_{i-1}}

\newcommand{\barz}{\overline{z}}

\begin{document}
\title[Extracting Kolmogorov complexity]{Extracting the Kolmogorov complexity of strings and sequences from sources with limited independence}

\author[Towson]{M. Zimand}{Marius Zimand}
\address[Towson]{Department of Computer and Information Sciences
  \newline Towson University}  
\email{mzimand@towson.edu}  
\urladdr{http://triton.towson.edu/\~{}mzimand}  
\thanks{The author is supported by NSF grant CCF 0634830.}


\begin{abstract} 
An infinite binary sequence has randomness rate at least $\sigma$ if, for almost every $n$, the Kolmogorov complexity of its prefix of length $n$ is at least $\sigma n$. It is known that for every rational $\sigma \in (0,1)$, on one hand,  there exists sequences with randomness rate $\sigma$  that can not  be effectively transformed into a sequence with randomness rate higher than $\sigma$ and, on the other hand, any two independent sequences with randomness rate $\sigma$ can be transformed into a sequence with randomness rate higher than $\sigma$. We show that the latter result holds even if the two input sequences have linear dependency (which, informally speaking, means that all prefixes of length $n$ of the two sequences have in common a constant fraction of their information).  The similar problem is studied for finite strings. It is shown that from any two strings with sufficiently large Kolmogorov complexity and sufficiently small dependence, one can effectively construct a string that is random even conditioned by any one of the input strings.
\end{abstract}

\maketitle

\section{Introduction}

The randomness rate of an object is the ratio between the information in the object and its length. An informal principle states that no reasonable transformation can guarantee an increase of the randomness rate. The principle has different instantiations depending on the meaning of ``object'', ``information,'' and ``reasonable transformation.'' For example, if $f$ is a mapping of the set of $n$-bit strings to the set of $m$-bit strings, then there is a distribution $X$ on the set of $n$-bit strings with Shannon entropy $n/2$ (\ie,  the randomness rate of $X$ is $1/2$) and the Shannon entropy of $f(X)$ is $\leq m/2$ (\ie, the randomness rate of $f(X)$ is $\leq 1/2$). Thus no transformation $f$ guarantees that its output has a randomness rate higher than that of its input. The case of infinite binary sequences (in short, sequences) is very interesting and has been recently the subject of intensive research. We say that a sequence $x$ has randomness rate at least $\sigma$ if $K(x \rest n) \geq \sigma \cdot n$, for all sufficiently large $n$. Here, $x \rest n$ is the prefix of length $n$ of $x$ and $K(\cdot)$ is the Kolmogorov complexity. A related notion is that of effective Hausdorff dimension of a sequence $x$, defined as: $\dimm(x) = \lim \inf K(x \rest n)/n$. Reiman and Terwijn~\cite{rei:t:thesis} have asked whether for any sequence $x$ with $\dimm(x) = 1/2$ there exists an effective transformation (formally, a Turing reduction) $f$ such that $\dimm(f(x)) > 1/2$. Initially, some partial negative results have been obtained for transformations $f$ with certain restrictions.  Reimann and Terwijn~\cite[Th 3.10]{rei:t:thesis}   have shown that the answer is NO if we require that $f$ is a many-one reduction.  This result has been extended by Nies and Reimann~\cite{nie-rei:c:wtt-Kolm-increase} to wtt-reductions.
Bienvenu, Doty, and Stephan~\cite{bie-dot-ste:c:haussdimension}  have obtained an impossibility result for the general case of Turing reductions, which, however, is valid only for \emph{uniform} transformations. More precisely, building on the result of Nies and Reimann, they have shown that for all constants $c_1$ and $c_2$, with $0 < c_1 < c_2 < 1$, there is no Turing reduction $f$ such that for any sequence $x$ with $\dimm(x) \geq c_1$  has the property that $\dimm(f(x)) \geq c_2$. In other words, loosely speaking, no effective uniform transformation is able to raise the randomness rate from $c_1$ to $c_2$. Finally, very recently, Miller~\cite{mill:t:KolmExtract} has fully solved the original question, by constructing a sequence $x$ with $\dimm(x) = 1/2$ such that, for any Turing reduction $f$, $\dimm(f(x)) \leq 1/2$ (or $f(x)$ does not exist). 

On the other hand, Zimand~\cite{zim:c:csr} has shown that it is possible to increase the randomness rate if the input consists of \emph{two sequences} that enjoy a certain type of \emph{independence}.   
Namely, we say that two sequences $x$ and $y$ are finitary-independent\footnote{In~\cite{zim:c:csr}, such sequences are called independent. The paper~\cite{cal-zim:c:dlt08} examines thoroughly the concept of algorithmic independence for sequences and introduces besides finitary-independence, a stronger concept which is called independence. We adopt here the terminology from~\cite{cal-zim:c:dlt08}.}  if for all $n$ and $m$, 
\begin{equation}
\label{e:fin-indep}
K(x \rest n ~y \rest m) \geq K(x \rest n) + K(y \rest m) - O(\max (\log n, \log m)).
\end{equation}
In~\cite{zim:c:csr}, it is shown that for any constant $0 < \tau \leq 1$, there is a Turing reduction $f$ such that, for any finitary-independent sequences $x$ and $y$, both with randomness rate $\geq \tau$, it holds that $f(x,y)$ has randomness rate arbitrarily close to $1$ (in particular, $\dimm(f(x,y)) = 1$). Moreover $f$ is a truth-table reduction and also $f$ is uniform in $\tau$.

To summarize, if we start with one source, it is impossible to effectively increase the randomness rate, while if we start with two finitary-independent sequences it is possible to increase the randomness rate to close to $1$ in a uniform and truth-table manner.

It is clear that the independence requirement plays an important role in the positive result. Since independence can be quantified, it is interesting to see what level of independence is needed for a positive result.

For a function $d: \nat \mapping \real$, we say that strings $u$ and $v$ have dependency $d$ if $K(u) + K(v) - K(uv)  \leq d(\max(|u|,|v|))$; we say that two sequences $x$ and $y$ have dependency $d$ if, for every $n$ and $m$ sufficiently large, the strings $x\rest n$ and $y \rest m$ have dependency $d$. With this terminology, sequences $x$ and $y$ are finitary-independent if they have dependency $c \cdot \log n$, for some positive constant $c$. 

The question becomes: How large can $d$ be so that an effective increase of the randomness rate is possible from two sequences with dependency $d$?
Miller's result shows that this is impossible for dependency $d(n) = n$, while the result in~\cite{zim:c:csr} shows that this is possible for dependency $d(n) = c \cdot \log n$. In fact, \cite{zim:c:csr} shows that, for certain combinations of parameters, an effective increase is possible even for dependency $d(n) = n^\alpha$, for some $0 < \alpha < 1$. More precisely, it is shown that for any $\tau > 0$ and $\delta > 0$, there exists $0 < \alpha < 1$ and a truth-table reduction $f$ such that for any sequences $x$ and $y$ that have randomness rate $\tau$ and dependency $d(n) = n^\alpha$, it holds that $f(x,y)$ has randomness rate $1 - \delta$.

In this paper, we improve the above result and show that one can effectively increase the randomness rate even for two  input sources that have linear dependency. More formally, our result is:
\begin{itemize}
\item[(1)] We show that for every $0< \tau \leq 1$ and $\delta > 0$, there exist $0< \alpha < 1$ and a truth-table reduction $f$ such that for any sequences $x$ and $y$ with randomness rate $\tau$ and dependency $d(n) = \alpha  n$, the sequence $f(x,y)$ has randomness rate $\geq (1 - \delta)$.

\end{itemize}

We also study the finite version of the problem, when the input consists of strings.  Similarly to the infinite case, our interest is in determining how many input strings and what level of dependency are necessary in order to exist an effective procedure that extracts Kolmogorov complexity. Vereshchagin and Vyugin~\cite[Th. 4]{ver-vyu:j:kolm} have shown that one input string is not enough. They construct a string $x$ so that any shorter string that has small Kolmogorov complexity conditioned by $x$ (in particular any string effectively constructed from $x$) has small Kolmogorov complexity unconditionally. On the other hand, Fortnow, Hitchcock, Pavan, Vinodchandran and Wang~\cite{fhpvw:c:extractKol} show that an input consisting of several independent strings can accomplish the task, when the number of strings in the input depends on the complexity of the strings. Formally, they show that, for any $\sigma$ there exists a constant $\ell$ and a polynomial-time procedure
that from an input consisting of $\ell$ $n$-bit strings $x_1, \ldots, x_\ell$, each with Kolmogorov complexity at least $\sigma n$, constructs
an $n$-bit string with Kolmogorov complexity $\succeq n - {\rm dep}(x_1, \ldots, x_\ell)$ (${\rm dep}(x_1, \ldots, x_\ell) = \sum_{i=1}^{\ell} K(x_i) - K(x_1 \ldots x_\ell)$ and $\succeq$ means that the inequality holds within an error of $O(\log n)$). In view of Vereshchagin-Vyugin result, the question is whether effective extraction of Kolmogorov complexity is possible from two input strings. 
We have two results in this regard:
\begin{itemize}
\item [(2)] We show that if strings $x$ and $y$ of length $n$ have dependency $\alpha n$ and complexity $\sigma n$, then it is possible to effectively construct a string of length $\approx 2\sigma \cdot n$ and complexity $\succeq (2\sigma - \alpha) \cdot n$, where $\approx$ ($\succeq$)  means that the equality (resp., the inequality) is within an error of $O(\log n)$. The construction is uniform in $x,y, \alpha, \sigma$. Note, however, that unlike the procedure from~\cite{fhpvw:c:extractKol}, the construction does not run in polynomial time.
\item[(3)] Our second result shows that from strings $x$ and $y$,  with sufficiently large complexity and sufficiently small dependency, it is possible to construct a string $z$ that has large complexity even conditioned by any of the input strings. More precisely if $x$ and $y$ are strings of length $n$ that have complexity $s(n)$ and dependency $\alpha(n)$, then it is possible to effectively construct a string of length $m \approx s(n)/2$ such that $K(z \mid x) \succeq m - \alpha(n)$ and 
$K(z \mid y) \succeq m - \alpha(n)$. The construction is uniform in $x,y, s(n), \alpha(n)$. This improves a result from~\cite{cal-zim:c:dlt08}, where the input consists of three strings $x_1, x_2, x_3$ and the construction produces a string $z$ with large $K(z \mid x_i)$, $i=1,2,3$.
\end{itemize}
Effective procedures that extract the Kolmogorov complexity of strings are related to randomness extractors. These are objects of major interest in computational complexity and there is a long and very active line of research dedicated to them. A randomness extractor is a procedure (which, ideally, runs in polynomial time) that improves the quality of a defective source of randomness. A source of randomness is modeled by a distribution $X$ on $\zon$, for some $n$, and its quality is modeled by the min-entropy of $X$ ($X$ has min-entropy $k$ if $2^{-k}$ is the largest probability that $X$ assigns to any string in $\zon$). The distribution $X$ is defective if its min-entropy is less than $n$, and is perfect if its min-entropy is equal to $n$, which implies that $X$ is the uniform distribution on $\zon$. In many applications, it is desirable to transform a defective distribution $X$ into a distribution $X'$ on a set of shorter strings which is close to the uniform distribution. Such a transformation is called a randomness extractor. Randomness extraction is not possible from a single source~\cite{san-vaz:j:quasirand}, but it is possible from two or more sources~\cite{vaz:j:sources}. Consequently, the research has focused on two types of extractors, seeded extractors and multi-source extractors. A seeded extractor extracts randomness from two independent distributions $X$ and $Y$, where $X$ is defective and defined on the set of $n$-bit strings  and $Y$ is perfect and defined on the set of $d$-bit strings, with $d$ much shorter than $n$ (typically $d = O(\log n)$). A $k$-multisource extractor takes as input $k$ defective distributions on the set of $n$-bit strings. For $k=2$, the best  multisource extractors  are (a) the extractor given by Raz~\cite{raz:c:multiextract} with one source having min-entropy $((1/2) + \alpha)n$ (for some small $\alpha$) and the second source having min-entropy polylog($n$), and (b) the extractor given by Bourgain~\cite{bou:j:multiextract} with both sources having min-entropy $((1/2) - \alpha)n$ (for some small $\alpha$). There is a clear analogy between randomness extractors and procedures that extract Kolmogorov complexity. In particular, the reader may compare results (2) and (3) discussed above  with existing $2$-multisource extractors, but we emphasize that there is a major difference in that extractors run in polynomial time, while the procedures in (2) and (3) are only in EXPSPACE. On the other hand, results (2) and (3) suggest that it might be possible to construct multisource extractors with sources having a certain level of dependence and/or with the output being random even conditioned by one of the sources.

A few words about the proof technique. At the highest level, our method follows the structure of the proofs in~\cite{zim:c:csr}. One key idea is taken from Fortnow et al.~\cite{fhpvw:c:extractKol}, who showed that a multisource extractor also extracts Kolmogorov complexity. Since multisource extractors with the parameters that are needed here are not known to exist, we construct a combinatorial object, called a balanced table, that is similar with a $2$-multisource extractor. A balanced table is a  $2$-dimensional $N \times N$ table with each entry having one of $M$ colors such that in each sufficiently large subrectangle all the colors appear approximately the same number of times (see Definition~\ref{d:balancedtable}). Using the probabilistic method, we show the existence of balanced tables with appropriate parameters. It follows that such tables can be effectively constructed using exhaustive search. Next, using arguments similiar to those in~\cite{fhpvw:c:extractKol}, we show that if $x$ and $y$ have sufficiently large  complexity and sufficiently small dependence, then the color of the entry in row $x$ and column $y$ of the table has large complexity. These ideas are sufficient to establish result~(2) (Theorem~\ref{t:result2}).  Results (1) (Theorem~\ref{t:sequences}) and (3) (Theorem~\ref{t:result3}) require non-trivial technical refinements of the basic method which are explained in the respective proofs.
\section{Preliminaries}
\label{s:prelim}
\subsection{Notation}
We work over the binary alphabet $\{0,1\}$. A string is an element of $\{0,1\}^*$ and a sequence is an element of $\{0,1\}^{\infty}$. If $x$ is a string, $|x|$ denotes its length. If $x$ is a string or a sequence and $n \in \nat$, $x \rest n$ denotes the prefix of $x$ of length $n$. The cardinality of a finite set $A$ is denoted $|A|$. For $n \in \nat$, $[n]$ denotes the set $\{1,2, \ldots, n\}$. Let $M$ be a standard Turing machine. For any string $x$, define the \emph{(plain) Kolmogorov complexity} of $x$ with respect to $M$, as 
\[K_M(x) = \min \{ |p| \mid M(p) = x \}.
\]
 There is a universal Turing machine $U$ such that for every machine $M$ there is a constant $c$ such that for all $x$,
\begin{equation}
\label{e:univ}
K_U(x) \leq K_M(x) + c.
\end{equation}
We fix such a universal machine $U$ and dropping the subscript, we let $K(x)$ denote the Kolmogorov complexity of $x$ with respect to $U$. For the concept of conditional Komogorov complexity, the underlying machine is a Turing machine that in addition to the read/work tape which in the initial state contains the input $p$, has a second tape containing initially a string $y$, which is called the conditioning information. Given such a machine $M$, we define the Kolmogorov complexity of $x$ conditioned by $y$ with respect to $M$ as 
\[K_M(x \mid y) = \min \{ |p| \mid M(p, y) = x \}.
\]
Similarly to the above, there exist  universal machines of this type and they satisfy the relation similar to Equation~\ref{e:univ}, but for conditional complexity. We fix such a universal machine $U$, and dropping the subscript $U$, we let $K(x \mid y)$ denote the Kolmogorov complexity of $x$ conditioned by $y$ with respect to $U$.

Let $\sigma \in [0,1]$. A sequence $x$ has randomness rate at least $\sigma$ if $K(x(1:n)) \geq \sigma \cdot n$, for almost every $n$ (\ie, the set of $n$'s violating the inequality is finite).

The procedures that we design for extracting the Kolmogorov complexity of strings or sequences are either computable functions (in the case of strings) or Turing reductions (in the case of sequences). In our result,  the Turing reduction is also uniform in two parameters $\tau$ and $\sigma$. Formally, such a Turing reduction   $f$ is represented by a two-oracle Turing machine $M_f$. The machine $M_f$ has access to two oracles $x$ and $y$, which are binary sequences. When $M_f$ makes the query ``$n$-th bit of first oracle?" (``$n$-th bit of second oracle?"), the machine obtains $x(n)$ (respectively, $y(n)$). On input $(\tau, \sigma, 1^n)$, where $\tau$ and $\sigma$ are  rational numbers  (given in some canonical representation), $M_f$ outputs one bit. We say that $f(x,y, \tau, \sigma,) = z \in \{0,1\}^\infty$, if for all $n$,  $M_f$ on input $(\tau,\sigma, 1^n)$ and working with oracles $x$ and $y$ halts and outputs $z(n)$.  
In case the machine $M_f$ halts on all inputs and with all oracles, we say that $f$ is a truth-table reduction.
\subsection{Limited Independence}
\label{s:indep}

\begin{definition} 
\label{d:indep}
\begin{itemize}
\item[(a)] The dependency of two strings $x$ and $y$ is ${\rm dep}(x,y) = K(x) + K(y) - K(xy)$.
\item[(b)] Let $d: \nat \mapping \real$. We say that strings $x$ and $y$ have dependency at most $d(n)$ if ${\rm dep}(x,y) \leq d(\max(|x|, |y|))$.
\item[(c)] Let $d: \nat \mapping \real$. We say that sequences $x$ and $y$ have dependency at most $d(n)$, if for every natural numbers $n$ and $m$, the strings $x \rest n$ and $y \rest m$ have dependency at most $d(n)$.
\end{itemize}
\end{definition}
\if
The definition says that, modulo additive logarithmic terms, there is no shorter way to describe the concatenation of any two initial segments of $x$ and $y$ than having the information that describes the initial segments.

It can be shown that the fact that $x$ and $y$ are independent is equivalent to saying that for every natural numbers $n$ and $m$,
\begin{equation}
K(x(1:n) \mid y(1:m)) \geq K(x(1:n)) - O(\log(n) + \log(m)).
\end{equation}
and
\begin{equation}
K(y(1:m) \mid x(1:n)) \geq K(y(1:m)) - O(\log(n) + \log(m)).
\end{equation}
Thus, if two sequences $x$ and $y$ are independent, no initial segment of one of the sequence can help in getting a shorter description of any initial segment of the other sequence, modulo additive logarithmical terms.
 \fi
\subsection{Balanced Tables}
\label{s:balanced}
Let $N$ and $M$ be positive integers. An $(N, M)$ table is a function $T:[N] \times [N] \mapping [M]$. It is convenient to view it as a two dimensional table with $N$ rows and $N$ columns where each entry has a color from the set $[M]$. If $B_1, B_2$ are subsets of $[N]$, the $B_1 \times B_2$ rectangle of table $T$ is the part of $T$ comprised of the rows in $B_1$ and the columns in $B_2$.
\begin{definition}
\label{d:balancedtable}
Let $T: [N] \times [N] \mapping [M]$ be an $(N,M)$ table and $S \leq N$ and $D \leq M$ be two positive integers. We say that the table is $(S, D)$-balanced if for every set $A \subseteq [M]$ with $|A| = M/D$ and for every sets $B_1 \subseteq [N], B_2 \subseteq [N]$ with $|B_1| \geq S, |B_2| \geq S$,
\[
| T^{-1}(A) \cap (B_1 \times B_2) | \leq 2 \cdot \frac{|A|}{M} \cdot |B_1 \times B_2|.
\]
\end{definition}
The above definition states that in any $B_1 \times B_2$ rectangle of $T$ and for any set $A$ of colors of size $M/D$, the fraction of occurences of colors in $A$ is bounded by $2 \cdot |A|/M$.
\begin{lemma}
\label{l-balancedtable}
Suppose $S^2 > 3M + 3M \ln D +6SD + 6SD \ln(N/S)$. Then there exists a table $T : [N] \times [N] \mapping [M]$ that is $(S,D)$-balanced.
\end{lemma}
\proof
The proof is by the probabilistic method. We color the $N$-by-$N$ table selecting for each entry independently at random a color from $[M]$. Let us fix $A \subseteq [M]$ with $|A| = M/D$, $B_1 \subseteq [N]$ with $|B_1| = S$ and $B_2 \subseteq [N]$ with $|B_2| = S$. Note that it is enough to prove the assertion for sets $B_1$ and $B_2$ of size exactly $S$.
By the Chernoff bounds,
\begin{equation}
\label{e:1}
{\rm Prob}\bigg( \frac{\mbox{number of $A$-colored cells in $B_1 \times B_2$}}{S^2} > 2 \frac{|A|}{M}\bigg) \leq e^{-(1/3) (|A|/M) S^2} = e^{-(1/(3D))S^2}.
\end{equation}
The number of possibilities of choosing the set $A$ as above is bounded by
\begin{equation}
\label{e:2}
{M \choose M/D} \leq (e\cdot D)^{M/D} = e^{M/D + (M/D) \ln D}.
\end{equation}
The number of possibilities of choosing the sets $B_1$ and $B_2$ as above is bounded by
\begin{equation}
\label{e:3}
{N \choose S}^2 \leq (e N/S)^{2S} = e^{2S + 2S \cdot \ln(N/S)}.
\end{equation}
The hypothesis ensures that the product of the upper bounds in Equations~(\ref{e:1}), ~(\ref{e:2}), and~(\ref{e:3}) is less than $1$. It follows from the union bound that there exists an $(S,D)$-balanced table.
\qed
\medskip

In our applications, $N$ and $M$ will be powers of two, $N=2^n$ , $M = 2^m$, and $[N]$ is identified with $\zo^n$ and $[M]$ is identified with $\zo^m$. We assume this setting in the following.
\begin{lemma} 
\label{l-bal-table-refined}
Let $T:  [N] \times [N] \mapping [M]$ be an $(S,M)$-balanced table. Let $v$ be a string with $|v| \leq m$. Then for all sets $B_1 \subseteq [N],  B_2 \subseteq [N]$ with $|B_1| \geq S, |B_2| \geq S$, the number of entries in the $B_1 \times B_2$ rectangle of $T$ that have a color whose prefix is $v$ is  $\leq 2 \cdot \frac{1}{2^{|v|}} \cdot |B_1 \times B_2|$.
\end{lemma}
\proof
First observe that, since the table is $(S,D)$-balanced with the value of the parameter $D$ equal to $M$, the definition of an $(S,D)$-balanced table implies that no color $a \in [M]$ occurs more than a fraction of $2/M$ times in any rectangle of $T$ with sizes $\geq S$. Let $v$ be a string of length of most $m$. Then $v$ has $2^{m-|v|}$ extensions of length $m$ and, as we have just noted, each such extension occurs at most a fraction $2/M$ in any rectangle with sizes $\geq S$. It follows that in any $B_1 \times B_2$ rectangle of $T$, all the extensions of $v$ taken together occur at most $2^{m-|v|} \cdot (2/M) \cdot |B_1 \times B_2| = (2/2^{|v|}) \cdot |B_1 \times B_2|$ times.
\qed
\section{Increasing the randomness rate of strings}
The next theorem shows that from two $n$-bit strings with complexity $\sigma n$ and dependency $\alpha n$, one can construct a string of length $\approx 2\sigma n$ and complexity $\approx ( 2\sigma - \alpha)n$.
\begin{theorem}
\label{t:result2}
For every $\sigma > 0$, for every $0 < \alpha < \sigma$, there is a computable function
$f:\zo^* \times \zo^* \mapping \zo^*$, that, for every $n$,  maps any pair of strings of length $n$ into a string of length $m = 2\sigma n - \log n$  and has the following property: for every sufficiently large $n$,
if $(x,y)$ is a pair of strings with
\begin{enumerate}
	\item $|x| = |y| = n$,
	\item $K(x) \geq \sigma n$, $K(y) \geq \sigma n$
	\item $(x,y)$ have dependency at most $\alpha n$,
\end{enumerate}
then
\[
K(f(x,y)) \geq (2\sigma - \alpha) n - 9 \log n.
\]
\end{theorem}
\proof
Let us fix $n$ and let $N = 2^n, m = 2\sigma n - \log n, M = 2^m, S= 2^{\sigma n}, d = \alpha n + 8 \log n$, and $D = 2^d$. Note that the requirements of Lemma~\ref{l-balancedtable} are satisfied and therefore there exists a table $T: [N] \times [N] \mapping [M]$ that is $(S,D)$-balanced. By brute force, we find the smallest (in some canonical sense) such table $T$. Note that the table $T$ can be described with $\log n + O(1)$ bits. We define $f(x,y)$ to be $T(x,y)$. Thus, let $z = T(x,y)$ for some strings $x$ and $y$ of length $n$ satisfying the requirements in the theorem hypothesis.  For the sake of obtaining a contradiction, suppose that $K(z) <  (2\sigma - \alpha) n - 9 \log n = m-d$.
Let $t_1 = K(x), t_2 = K(y)$. From the properties of $x$ and $y$, $t_1 \geq \sigma n$ and $t_2 \geq \sigma n$. Let $B_1 = \{u \in \zo^n \mid K(u) \leq t_1 \}$, $B_2 = \{v \in \zo^n  \mid K(v) \leq t_2\}$ and
$A = \{w \in \zo^m \mid K(w) < m-d\}$. We have $|B_1| \leq 2^{t_1 + 1}$, $|B_2| \leq 2^{t_2 + 1}$ and $|A| < 2^{m-d}$. We take $B_1'$ and $B_2'$ with $|B_1'| = 2^{t_1 + 1}, |B_2'| = 2^{t_2 + 1}$, $B_1 \subseteq B_1'$ and $B_2 \subseteq B_2'$. Since the table $T$ is $(S,D)$-balanced,
\[
\begin{array}{ll}
|T^{-1}(A) \cap (B_1 \times B_2)| \leq |T^{-1}(A) \cap (B_1' \times B_2')| & \leq 2 \cdot \frac{|A|}{M} \cdot |B_1' \times B_2'| \\
& \leq 2 \cdot 2^{m-d} \frac{1}{2^m} \cdot 2^{t_1 + 1} \cdot 2^{t_2 + 1} \\
& \leq 2^{t_1 + t_2 - d + 3}. 
\end{array}
\]
Note that $(x,y) \in T^{-1}(A) \cap (B_1 \times B_2)$ and that $T^{-1}(A) \cap (B_1 \times B_2)$ can be enumerated if we are given $t_1, t_2$ and $n$ (from which $(m-d)$ and a description of table $T$ can be determined). Therefore $xy$ can be described by the rank of $(x,y)$ in the above enumeration and by information needed for performing that enumeration. Thus
\[
\begin{array}{ll}
K(xy) & \leq t_1 + t_2 - d + 2 (\log t_1 + \log t_2 + \log n ) + O(1) \\
& \leq t_1 + t_2 - d + 7 \log n.
\end{array}
\]
For the second inequality, we took into consideration that $t_1 \leq n$ and $t_2 \leq n$. On the other hand, since $x$ and $y$ have dependency bounded by $\alpha n$.
\[
K(xy) \geq t_1 + t_2 - \alpha n.
\]
Keeping in mind that $d = \alpha n + 8 \log n$, we have obtained a contradiction.
\qed

The next theorem shows from two $n$-bit strings with complexity $s(n)$ and dependency $\alpha(n)$, one can construct a string of length $m \approx s(n)/2$ with complexity conditioned by any one of the input strings $\approx m -\alpha(n)$.
\begin{theorem}
\label{t:result3}
For every computable function $s(n)$ verifying $6 \log n < s(n) \leq n$ and every function $\alpha(n)$, there is a computable function $f: \zo^* \times \zo^* \mapping \zo^*$ that, for every $n$, maps any pair of strings of length $n$ into a string of length $m = s(n)/2 - 7 \log n$ and has the following property: for every sufficiently large $n$, if $(x,y)$ is a pair of strings with
\begin{enumerate}
	\item $|x| = |y| = n$,
	\item $K(x) \geq s(n), K(y) \geq s(n)$,
	\item $(x,y)$ has dependency at most $\alpha(n)$
\end{enumerate}
then
\[
\begin{array}{ll}
K(f(x,y) \mid x) & \geq m - \alpha(n) - 11 \log n, \\
K(f(x,y) \mid y) & \geq m - \alpha(n) - 11 \log n. 
\end{array}
\]
\end{theorem}
\proof
We fix $n$ and let $N=2^n, m = s(n)/2 - 7 \log n, M= 2^m, S= 2^{s(n)/2}, D=M, t= \alpha(n) + 11 \log n$. The requirements of Lemma~\ref{l-balancedtable} are satisfied and therefore there exists a table $T: [N] \times [N] \mapping [M]$ that is $(S,D)$-balanced. By brute force, we find the smallest (in some canonical sense) such table $T$. The table $T$ is determined by $n$ and $s(n)$, and, thus, can be described with $\log n + \log s(n) + O(1)$ bits. Note that, since $D=M$, it holds that for every color $a \in [M]$ and for every subsets $B_1 \subseteq [N], B_2 \subseteq [N]$ with
$|B_1| \geq S, |B_2| \geq S$, the number of occurrences of $a$ in the $B_1 \times B_2$ subrectangle of $T$ is bounded by $(2/M) \cdot |B_1 \times B_2|$.

We define $f(x,y)$ to be $T(x,y)$. Thus, let $z = T(x,y)$ for some strings $x$ and $y$ of length $n$ satisfying the requirements in the theorem hypothesis. We need to show that $K(z \mid x)$ and $K(z \mid y)$ are at least $m - \alpha(n) - 11 \log n$. We show this relation for $K(z \mid y)$ (the proof for $K(z \mid x)$ is similar). For the sake of obtaining a contradiction, suppose that $K(z \mid y) <  m - \alpha(n) - 11 \log n = m - t$. Let $t_1 = K(x)$. Note that $t_1 \geq s(n)$. Let $B = \{u \in \zo^n \mid K(u) \leq t_1\}$. Note that $2^{t_1 + 1} > |B| \geq 2^{s(n)/2} = S$. We say that a column $u \in [N]$ is \emph{bad for color $a \in [M]$ and $B$} if
the number of occurrences of $a$ in the $B \times \{u\}$ subrectangle of $T$ is greater that $(2/M) \cdot |B|$ and we say that $u$ is \emph{bad for $B$} if it is bad for some color $a$ and $B$. For every $a \in [M]$, the number of $u$'s that are bad for $a$ and $B$ is $< S$ (because $T$ is $(S,D)$-balanced). Therefore, the number of $u$'s that are bad for $B$ is $< M \cdot S$. Given $t_1$ and a description of the table $T$, one can enumerate the set of $u$'s that are bad for $B$. This implies that any $u$ that is bad for $B$ can be described by its rank in this enumeration and the information needed to perform the enumeration. Therefore, if $u$ is bad for $B$,
\[
\begin{array}{ll}
K(u) & \leq \log (M \cdot S) + 2 (\log t_1 + \log n + \log s(n))+O(1) \\
& \leq m + s(n)/2 + 6\log n +O(1) \\
& < s(n),
\end{array}
\]
provided $n$ is large enough.
Since $K(y) \geq s(n)$, it follows that $y$ is good for $B$.

Let $A = \{w \in [M] \mid K(w \mid y) < m- t\}$.  We have $|A| < 2^{m-t}$ and, by our assumption, $z \in A$. Let $G$ be the subset of $B$ of positions in the strip $B \times \{y\}$ of $T$ having a color from $A$ (formally, $G= {\rm proj}_1 (T^{-1}(A) \cap (B \times \{y\})$) . Note that $x$ is in $G$. Each color $a$ occurs in the strip $B \times \{y\}$ at most $(2/M) \cdot |B|$ (because $y$ is good for $B$). Therefore the size of $G$ is bounded
by 
\[
|A| \cdot(2/M) \cdot |B| \leq 2^{m-t} \cdot(2/M) \cdot 2^{t_1+1} < 2^{t_1 - t + 2}.
\]
Given $y, t_1, m-t$ and a description of the table $T$, one can enumerate the set $G$. Therefore, $x$ can be described by its rank in this enumeration and by the information needed to perform the enumeration. It follows that
\[
\begin{array}{ll}
K(x \mid y) & \leq t_1 - t + 2 + 2(\log t_1 + \log(m-t) + \log n + \log s(n)) + O(1) \\
& \leq t_1 - t + 8 \log n +O(1) \\
& = t_1 - \alpha(n) - 3 \log n + O(1)\\
& = K(x) - \alpha(n) - 3 \log n +O(1).
\end{array}
\]
Since $K(xy) \leq K(y) + K(x \mid y) + 2 \log n + O(1)$ (this holds for every $n$-bit strings $x$ and $y$), we obtain
\[
\begin{array}{ll}
K(x y) & \leq K(y) + K(x) - \alpha(n) - 3 \log n + 2 \log n + O(1) \\
& \leq K(y) + K(x) - \alpha(n) - \log n + O(1),
\end{array}
\]
which contradicts that $x$ and $y$ have dependency at most $\alpha(n)$.
\qed

\section{Increasing the randomness rate of sequences}
We prove that the randomness rate of sequences can be effectively increased even from two sequences having linear dependence.
\begin{theorem}
\label{t:sequences}
There exists a truth-table reduction $f$ with the following property. For any rational numbers $\tau > 0$ and $\delta >0$, there exists $\alpha > 0$ such that for any sequences $x$ and $y$ with randomness rate at least $\tau$ and dependency at most $\alpha n$, $f(x,y, \tau, \delta)$ has randomness rate at least $1-\delta$. Moreover, the reduction $f$ is uniform in $x,y, \tau$ and $\delta$.
\end{theorem}
\proof~The plan is as follows. We split $x$ into strings $x_1 x_2 \ldots x_i \ldots$ and $y$ into strings $y_1 y_2 \ldots y_i \ldots$. For each $i$, let $\barx_{i} = x_1 \ldots x_i$ and $\bary_{i} = y_1 \ldots y_i$. The splitting is done in such a way that $x_i$ and $y_i$ have complexity close to $\tau |x_i|$ and respectively close to $\tau |y_i|$ even conditioned by $\barx_{i-1} \bary_{i-1}$. Next, for each $i$, we construct a balanced table $T_i$ with appropriate parameters and take $z_i = T(x_i, y_i)$. The output of the truth-table reduction is the sequence $z = z_1 z_2 \ldots z_i \ldots$. As in the case of strings, it follows that $z_i$ has high complexity and actually this holds even conditioned by $\barz_{i-1} = z_1 z_2 \ldots z_{i-1}$. So far, the proof is as in~\cite{zim:c:csr}. The point of departure is that in order for the construction to work with inputs having linear dependence, we need to take the length of $z_i$ exponential in $i$ (rather than quadratic in $i$, which was the case in~\cite{zim:c:csr}). This creates difficulties in showing that every ``intermediate'' prefix of $z$ (\ie, a string that is an extension of $\barz_{i-1}$ and a prefix of $\barz_i$, for some $i$) has high complexity. To handle this, we argue that even prefixes of $z_i$ have relatively high Kolmogorov complexity conditioned by $\barx_{i-1} \bary_{i-1}$ (see Lemma~\ref{l:block}) and then the argument for ``intermediate'' strings  forks into two cases depending on whether the string is long or short (see Lemma~\ref{l:combination}).

We proceed with the formal proof.

We fix rational numbers $\tau > 0$ and $\delta > 0$.
Let $x$ and $y$ be sequences with randomness rate at least $\tau$.
Let $\epsilon = \delta/4$. 

We split $x = x_1 x_2 \ldots x_i \ldots$ and $y = y_1 y_2 \ldots y_i \ldots$ and let $n_i = |x_i| = |y_i|$.
We'll take $n_i = B^i$ for some constant $B$, given by the next lemma.
\begin{lemma}
\label{l:B}
There exists a constant $B > 1$ with the following properties:
\begin{itemize}
	\item[(a)] For every $i$, $K(x_i \mid \barx_{i-1} \bary_{i-1}) \geq 0.99 \tau n_i$ and $K(y_i \mid \barx_{i-1} \bary_{i-1}) \geq 0.99 \tau n_i$.
	\item[(b)] For any $\alpha > 0$, if $(x,y)$ have dependency $\alpha n$,
then, for all $i$
\[
K(x_iy_i \mid \barx_{i-1}\bary_{i-1}) \geq K(x_i \mid \barx_{i-1}\bary_{i-1}) + K(y_i \mid \barx_{i-1}\bary_{i-1}) - (2.1) \cdot \alpha \cdot n_i.
\]
\end{itemize}
\end{lemma}
\proof (Sketch.) The proof is similar to an analogous result from~\cite{zim:c:csr}.  For (a), it is easy to show that $B$ can be taken  large enough so that the length of $x_i$ is so much larger than the length of $\barxy$ that the complexity of $x_i$ does not decrease too much if it is conditioned by $\barxy$.

The proof of (b) passes through the following intermediate steps:

(1) We show that for $B$, $i$ and $j$ sufficiently large, $$K(\bary_i \barx_j) = K(\bary_i) + K(\barx_j) \pm 1.001 \alpha (B^i + B^j).$$ (This is the analogue of Lemma 4.5 from~\cite{zim:c:csr}).

(2) We show that $B$, $i$ and $j$ sufficiently large, $$K(x_i \mid \barx_{i-1} \bary_j) = K(x_i \mid \barx_{i-1}) \pm 2.004 \alpha (B^i + B^j).$$ (This is the analogue of Lemma 4.6 from~\cite{zim:c:csr}; the constants are not optimized).

Next, the statement can be shown similarly to Lemma 4.7 from~\cite{zim:c:csr}).
\qed

For the rest of this section, we fix the following parameters as follows:
\begin{itemize}
\item The constant $B$ is as given by Lemma~\ref{l:B},
\item $\alpha = (1/3) \epsilon^2 \cdot(0.97\tau) \cdot (1/B)$.

\item For each $i$, 
$N_i = 2^{n_i}$,
$S_i = 2^{(0.98\tau)\cdot n_i}$,
$m_i = (0.97 \tau) \cdot n_i$,
$M_i = 2^{m_i}$
$D_i = M_i$.
\end{itemize}
The parameters satisfy the requirements of Lemma~\ref{l-balancedtable} and, thus, for each $i$, there exists a table $T_i : [N_i] \times [N_i] \mapping[M_i]$ that is $(S_i, D_i)$-balanced. For every $i$, given $i$, a smallest (in some canonical sense) such table $T_i$ can be constructed by exhaustive search. We fix these tables $T_i$ and define $z_i = T_i(x_i, y_i)$ and next $z = z_1 z_2 \ldots z_i \ldots$. Clearly $z$ is constructed by a truth-table reduction $f$ from input sequences $x$ and $y$. We will show that $z$ has randomness rate at least $1 - \delta$.
\begin{lemma}
\label{l:block}
For every $i$ sufficiently large, each prefix $v$ of $z_i$ has $K(v \mid \barxy) \geq |v| - 3\alpha \cdot n_i$.
\end{lemma}
\proof
Suppose that there is a prefix $v$ of $z_i$ with $K(v \mid \barxy) < |v| - 3\alpha \cdot n_i$. We define:
\begin{itemize}
	\item $t_1 = K(x_i \mid \barxy)$, $t_2 = K(y_i \mid \barxy)$,
	\item $B_1 = \{u \in \zo^{n_i} \mid K(u \mid \barxy) \leq t_1 \}$, 
	
	\item $B_2 = \{u \in \zo^{n_i} \mid K(u \mid \barxy) \leq t_2 \}$,
	\item $A= \{w \in \zo^{|v|} \mid K(w \mid \barxy) < |v| - 3 \alpha \cdot n_i\}$.
\end{itemize}
Note that $t_1 \geq 0.99 \tau \cdot n_i$, $t_2 \geq 0.99 \tau \cdot n_i$ (by Lemma~\ref{l:B}), $2^{t_1 + 1} > |B_1| \geq 2^{0.98 \tau \cdot n_i} = S_i$, $2^{t_2 + 1} > |B_2| \geq 2^{0.98 \tau \cdot n_i} = S_i$ and $|A| < 2^{|v| - 3\alpha n_i}$. Let $G$ be the set of entries (represented by their coordinates in the table) in the $B_1 \times B_2$ rectangle
of the table $T_i$ that have a color with a prefix in $A$. By Lemma~\ref{l-bal-table-refined}, the cardinality of $G$  is at most
\[
\begin{array}{ll}
2 \cdot \frac{|A|}{2^{|v|}} \cdot |B_1 \times B_2| & \leq 2 \cdot 2^{|v| - 3\alpha n_i} \cdot \frac{1}{2^{|v|}} \cdot 2^{t_1 + 1} \cdot 2^{t_2 + 1} \\
& = 2^{t_1 + t_2 - 3 \alpha n_i +3}.
\end{array}
\]
Note that $(x_i, y_i)$ belongs to $G$ and that $G$ can be enumerated given $\barxy$, $t_1, t_2$, and $|v| - 3\alpha \cdot n_i$ (observe that $i$ can be determined from $\barxy$ and thus the table $T_i$ can be constructed). Therefore $x_i y_i$ can be described by its rank in the enumeration of $G$ and by the information needed to perform this enumeration. This implies 
\[\begin{array}{ll}
K(x_i y_i \mid \barxy) & \leq t_1 + t_2 - 3\alpha \cdot n_i + 2(\log t_1 + \log t_2 + \log(|v| - 3\alpha n_i)) + O(1) \\
& \leq t_1 + t_2 - 3\alpha \cdot n_i + O(\log n_i).
\end{array}
\]
On the other hand, by Lemma~\ref{l:B},
\[
K(x_i y_i \mid \barxy) \geq K(x_i \mid \barxy) + K(y_i \mid \barxy) - (2.1) \cdot \alpha \cdot n_i.
\]
If $i$ is large, the last two inequalities conflict each other and we obtain a contradiction.
\qed
\smallskip

The next lemma finishes the proof of Theorem~\ref{t:sequences}.
\begin{lemma}
\label{l:combination}
For each sufficiently long prefix $w$ of $z$, $K(w) \geq (1-4\epsilon) |w|$.
\end{lemma}
\proof
For some $i$, the prefix $w$ is of the form $w = z_1 \ldots z_{i-1} v_i$, with $v_i$ a prefix of $z_i$.
Let $\gamma = (1/\epsilon) \cdot (3\alpha)$. We consider two cases:

{\bf Case 1: $v_i$ is long.} Suppose $|v_i| \geq \gamma \cdot n_i$.

Then $K(v_i \mid \barx_{i-1}\bary_{i-1}) \geq |v_i| - 3\alpha \cdot n_i \geq |v_i| - (3\alpha/\gamma) \cdot |v_i| = (1-\epsilon) |v_i|$.
This implies $K(v_i \mid z_1 \ldots z_{i-1}) > (1-\epsilon) \cdot |v_i| - O(1) \geq (1-2\epsilon)|v_i|$, because each $z_j$ can be constructed from $x_j$ and $y_j$.
By induction, it follows that $K(z_1 z_2 \ldots z_{i-1} v_i) \geq (1-3\epsilon) |z_1 z_2 \ldots z_{i-1} v_i|$. For the induction step, the argument goes as follows:
\[
\begin{array}{ll}
K(z_1 z_2 \ldots z_{i-1} v_i) & \geq K(z_1 \ldots z_{i-1}) + K(v_i \mid z_1 \ldots z_{i-1}) \\
& \quad \quad  - O(\log (m_1 + \ldots + m_{i-1}) + \log(|v_i|)) \\
& \geq (1-3\epsilon)(m_1 + \ldots + m_{i-1}) + (1-2\epsilon)|v_i| \\
& \quad \quad - O(\log(m_1 + \ldots +m_{i-1}) + \log |v_i|)) \\
& > (1-3\epsilon) (m_1 + \ldots + m_{i-1} + |v_i|).
\end{array}
\]
In the last step, we have used the fact that $\log(m_1 + \ldots + m_{i-1}) = O(i), \log|v_i| = O(i)$ and $|v_i| = \Omega(B^i)$.

{\bf Case 2: $v_i$ is short.} Suppose $|v_i| < \gamma \cdot n_i$.

For a contradiction, suppose $K(z_1 z_2 \ldots z_{i-1} v_i) < (1-4\epsilon) |z_1 z_2 \ldots z_{i-1} v_i|$.
Note that $z_1 z_2 \ldots z_{i-1}$ can be reconstructed from a descriptor of $z_1 z_2 \ldots z_{i-1} v_i$.
This implies 
\[
\begin{array}{ll}
K(z_1 z_2 \ldots z_{i-1})  & < (1-4\epsilon)(m_1 + m_2 +\ldots + m_{i-1} + |v_i|) +O(1) \\
& =  (1-4 \epsilon) (m_1 + \ldots +m_{i-1}) + (1-4\epsilon)|v_i| + O(1) \\
& \leq (1-4 \epsilon) (m_1 + \ldots +m_{i-1}) + (1-4\epsilon)\gamma \cdot n_i \\
& \leq (1-4 \epsilon) (m_1 + \ldots +m_{i-1}) + (1-4\epsilon) \cdot (1/\epsilon) (3 \alpha) \cdot n_i.
\end{array}
\]
But the second term is less than $\epsilon (m_1 + \ldots + m_{i-1})$ (due to the choice of $\alpha$). This implies that $K(z_1 z_2 \ldots z_{i-1}) \leq (1-3\epsilon) (m_1 + m_2 + \ldots + m_{i-1})$, which, by Case 1, is not possible.
\qed
\qed
\smallskip

{\bf Note.} It remains an open issue whether from input sequences $x$ and $y$ (even independent) one can construct a sequence $z$ that has high randomness rate conditioned by any one of the input sequences. In other words, the infinite analogue of Theorem~\ref{t:result3} is open.

\newcommand{\etalchar}[1]{$^{#1}$}


\end{document}